\documentclass[jkps,preprint,fleqn,showkeys]{revtex4}
\usepackage[pdftex]{graphicx}
\usepackage{amssymb}
\usepackage{amsmath}
\usepackage{bm}
\usepackage{xcolor}
\begin{document}
\setcounter{page}{0}
\title[]{Optimizing Sterile Neutrino Searches: Impact of Position Resolution in Short-Baseline Reactor Experiments}
\author{Young Ju \surname{Ko}}
\email{yjkophys@jejunu.ac.kr}
\affiliation{Department of Physics, Jeju National University, Jeju 63243, Korea}

\date[]{Received \today}

\begin{abstract}
The impact of position resolution on the sensitivity of short-baseline reactor neutrino experiments searching for light sterile neutrinos is investigated. Detailed simulations are conducted to evaluate two detector configurations: a segmented detector and an opaque liquid scintillator (OLS) detector, each positioned at two candidate research reactor sites, HANARO and Kijang. For both detector types, pseudo-data are generated under realistic assumptions regarding neutrino flux, detector response, and background levels. Oscillation analyses are performed to estimate detector sensitivity, incorporating both statistical and systematic uncertainties. The results indicate that the OLS detector, owing to its superior position resolution, achieves leading sensitivity across a wide range of oscillation parameters, even under conservative experimental conditions. These findings underscore the potential of OLS technology as a highly effective approach for future sterile neutrino searches.
\end{abstract}


\maketitle

\section{Introduction}
In 2011, the reactor antineutrino anomaly (RAA) was reported~\cite{PhysRevD.83.073006}, showing consistency with neutrino anomaly observed in gallium experiments~\cite{ANSELMANN1995440,PhysRevC.73.045805}. The RAA can be interpreted as the oscillation of light sterile neutrinos at the eV scale, which reignited interest in this possibility. This has led to numerous proposals for very short-baseline reactor neutrino experiments. Several of these proposals have been realized, with experiments either conducted or still ongoing~\cite{PhysRevLett.118.121802,nature.613.257,PhysRevD.103.032001,ALEKSEEV201856,Abreu_2021}. However, except for Neutrino-4~\cite{Serebrov2019}, no definitive positive evidence has been observed.

Recently, the BEST experiment provided additional support for the gallium anomaly~\cite{PhysRevLett.128.232501}, interpreted as a sign of sterile neutrino oscillations. Furthermore, the IceCube collaboration reported a weak but potentially indicative signal related to sterile neutrinos~\cite{PhysRevLett.133.201804}. The current experimental landscape presents a mix of positive and negative results, leaving the field ambiguous. To resolve this, further experiments are underway or proposed.

New detector technologies are being explored to enhance results from very short-baseline reactor neutrino experiments. Traditionally, experiments used homogeneous or segmented scintillator-based detectors, each with specific advantages and disadvantages. In particular, the segmented detectors have the advantage of being free from the use of a reference spectrum, which reduces systematic uncertainty when the baseline is sufficiently short. Therefore, segmented detectors may be preferred for research reactor experiments that can be installed at closer distances than experiments using commercial reactors.

A novel design using opaque media~\cite{Cabrera2021} offers excellent position resolution, introducing more pronounced trade-offs. In addition to the oscillation analysis advantages offered by segmented detectors, particle identification, which can distinguish between beta and gamma-ray signals, can lead to a dramatic reduction in background. This is a great advantage for experiments at research reactors with relatively high background due to shallow overburden.

However, it may have limited energy resolution due to a short light attenuation length. Of course, there is room for improvement in energy resolution depending on the opacity of the media or the gap between fibers, but this study considers the energy resolution to be somewhat worse than other types of detectors. Therefore, the key question is whether enhanced position resolution outweighs the reduced energy resolution. This study provides a preliminary feasibility assessment, estimating detector sensitivity to sterile neutrino searches as a function of various experimental parameters.

\section{Estimation of Sensitivities}
\subsection{Assumptions for Experimental Setup}
The NEOS experiment, conducted at the Hanbit Nuclear Power Plant in South Korea, deployed a neutrino detector at Unit 5, securing a baseline of approximately 24 meters. With a sufficiently high thermal power of 3 gigawatts and the presence of substantial overburden, a homogeneous type detector was adopted. However, no significant signals indicative of sterile neutrino oscillations were observed~\cite{PhysRevLett.118.121802}. The reliance on reactor neutrino flux models, an inherent limitation of single-detector setups, largely negates the statistical advantage gained from increased data collection. Consequently, experiments under similar conditions offer limited benefits even if conducted over longer durations or with larger detector volumes.

In contrast, experiments like STEREO and PROSPECT operate at research reactors, which have relatively lower thermal power and less overburden, resulting in a lower signal-to-background ratio. However, the advantage of closer proximity to the reactor provides significant benefits. To maximize these benefits and enhance background reduction, a segmented type detector was employed. This configuration can effectively function as a set of multiple identical and independent detectors placed at different distances from the reactor core, similar to the far-to-near ratio technique employed in multi-detector experiments~\cite{nature.613.257,PhysRevD.103.032001}.

In the environment of research reactors, the primary advantage of opaque liquid scintillator (OLS) can be further maximized~\cite{Cabrera2021}. OLS's exceptional position resolution, similar to that of segmented ones, has a greater impact on oscillation analyses as the baseline is shortened. When the baseline reaches the meter scale, improvements in position resolution for a meter-scale detector can significantly enhance sensitivity. Additionally, this approach could incorporate the benefits of a multi-detector setup. The ability to mitigate flux model dependence in multidetector configurations—typically achieved by comparing event rates at different baselines—can instead be realized within a single detector by analyzing events at different positions inside it.

Furthermore, superior position resolution plays a crucial role in background suppression. OLS technology offers a resolution high enough to distinguish interactions caused by beta particles and gamma rays~\cite{Cabrera2021}. If the prompt signal by the positron from inverse beta decay (IBD) can be effectively separated from gamma-induced background events, most expected background sources could be eliminated. In very short-baseline research reactor experiments, the relatively low thermal power results in a smaller number of IBD events. Consequently, the high background caused by the low overburden becomes an even more critical issue, potentially limiting the sensitivity of the experiment. The use of OLS can significantly mitigate this issue.

Based on these considerations, this study estimates the sensitivities for two detector configurations at two potential experimental sites. The first candidate site is the HANARO research reactor, a 30-MW thermal power facility located at the Korea Atomic Energy Research Institute in Daejeon. The second is a 15MW research reactor currently under construction in Kijang, Busan. The assumed baselines are 12 meters for HANARO and 7 meters for the Kijang reactor, determined based on expert consultation on reactor size and surrounding infrastructure. The background rate is assumed to be identical for both sites.

Two detector configurations are considered: a segmented detector, similar to STEREO, and a homogeneous detector that utilizes OLS. It is assumed that both configurations have a rectangular volume of $1\times2\times1\,\mathrm{m}^3$, positioned so that the variation in the distance from the reactor core is 2 meters. In the segmented detector type, five detector modules of 40 cm each are assumed, with position resolution effects not explicitly considered for each module. In contrast, for the OLS detector, a position resolution of 5 cm is assumed in each of the x, y, and z directions, based on reports suggesting centimeter-level resolution. 

Furthermore, the reactor core is considered a volume $50\times50\times50\,\mathrm{cm}^3$, with randomized neutrino generation points within it. Given the size of the reactor, I do not think that there is much merit in having smaller modules in the segmented detector. The assumed energy resolution is $5\%/\sqrt{E/\mathrm{MeV}}$ and $7\%/\sqrt{E/\mathrm{MeV}}$ for the segmented and OLS detector, respectively, and the Huber's flux model for $^{235}$U~\cite{PhysRevC.84.024617} and the Vogel \& Beacom's calculation for the IBD cross section~\cite{PhysRevD.60.053003} are used to generate neutrino energy.

The background rates for PROSPECT and STEREO were reported to be 530 and 440 per day~\cite{nature.613.257,PhysRevD.103.032001}, respectively, and a similar level of 500 per day was assumed for the segmented detector in this study. For the OLS detector, although further research is needed on the relevant part, considering that gamma-ray and positron signals can be identified, it is expected that there will be almost no background, and it is assumed to be 1\% compared to the segmented detector. A function $1/E^2$ is used for the background energy distribution.

\subsection{Sensitivity of Segmented Detector}
The segmented detector can effectively function as a set of multiple identical and independent detectors placed at different distances from the reactor core. This configuration allows the use of spectral or rate ratios between detector segments, similar to the far-to-near ratio technique employed in multi-detector experiments. Such ratios serve to cancel out correlated systematic uncertainties, leading to significantly improved experimental sensitivity. Among these uncertainties, the reactor antineutrino flux model stands out as the dominant source. The ability to eliminate dependence on this model is an important advantage.

To estimate expected detector performance and evaluate sensitivity, a large number of events must be generated to minimize statistical fluctuations. Each event is assigned two key pieces of information: the production position within the reactor core and the interaction position inside the detector. The production point is randomly selected within the assumed reactor volume. The baseline, defined as the distance between the production and detection points, is then sampled uniformly. For each generated baseline, a random detection point is chosen on the surface of a sphere centered at the production point with the sampled radius. Only events in which this interaction point lies within the detector volume are retained. This process is repeated until a total of $10^8$ valid events are recorded for each experimental site.

\begin{figure}[t]
  \centering
  \includegraphics[width=0.32\linewidth]{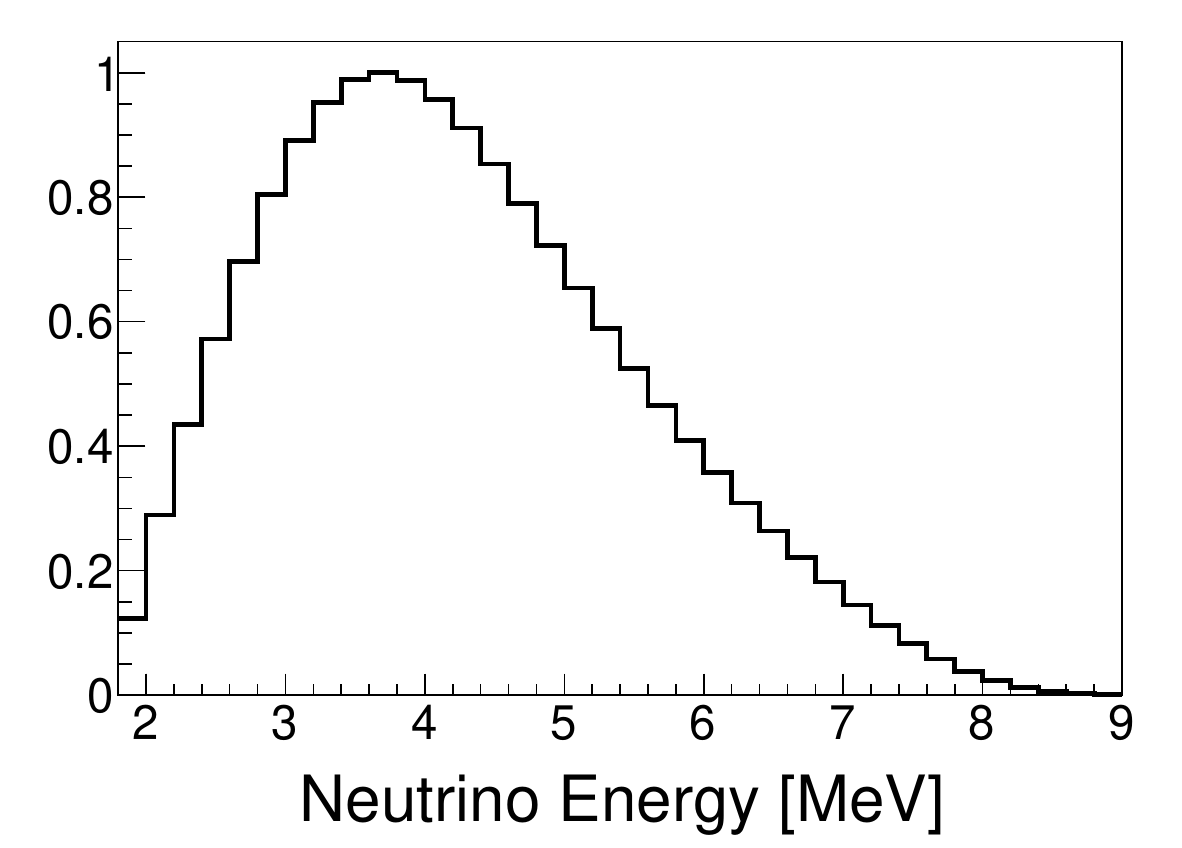}
  \includegraphics[width=0.32\linewidth]{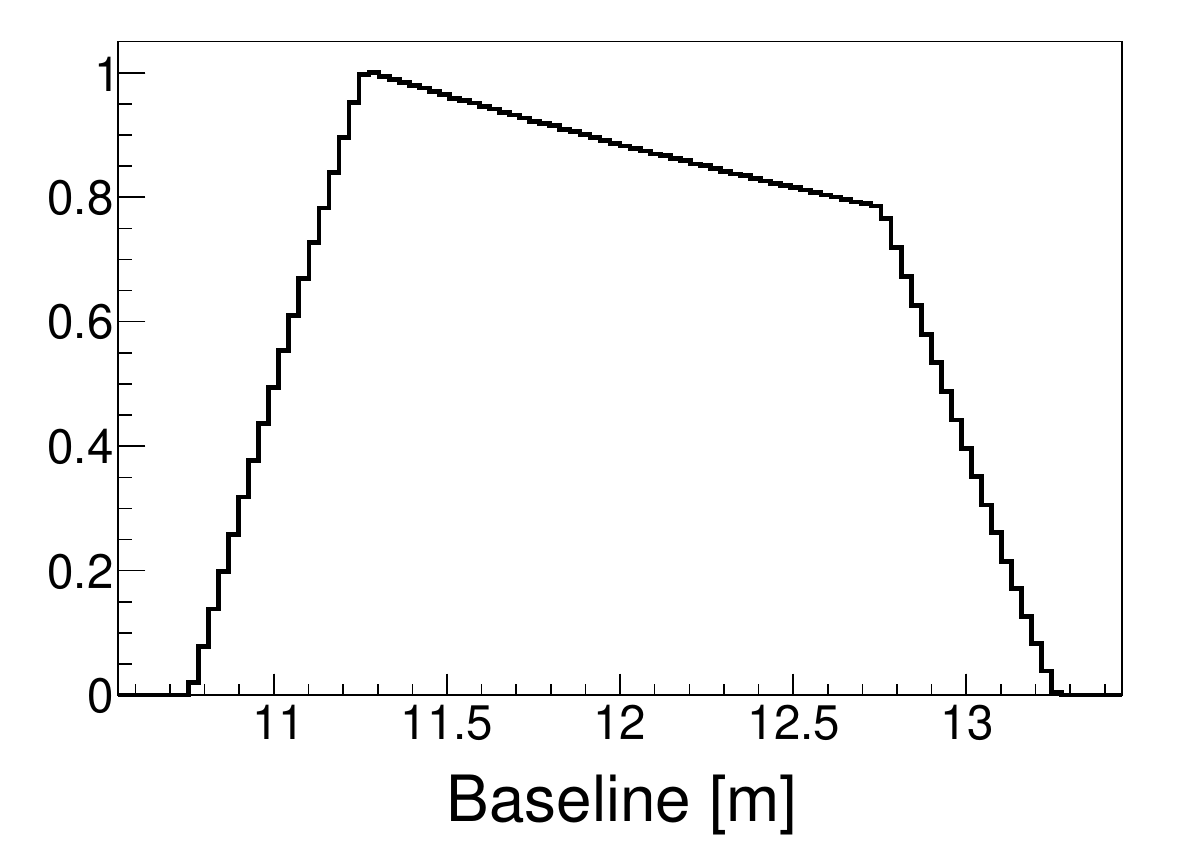}
  \includegraphics[width=0.32\linewidth]{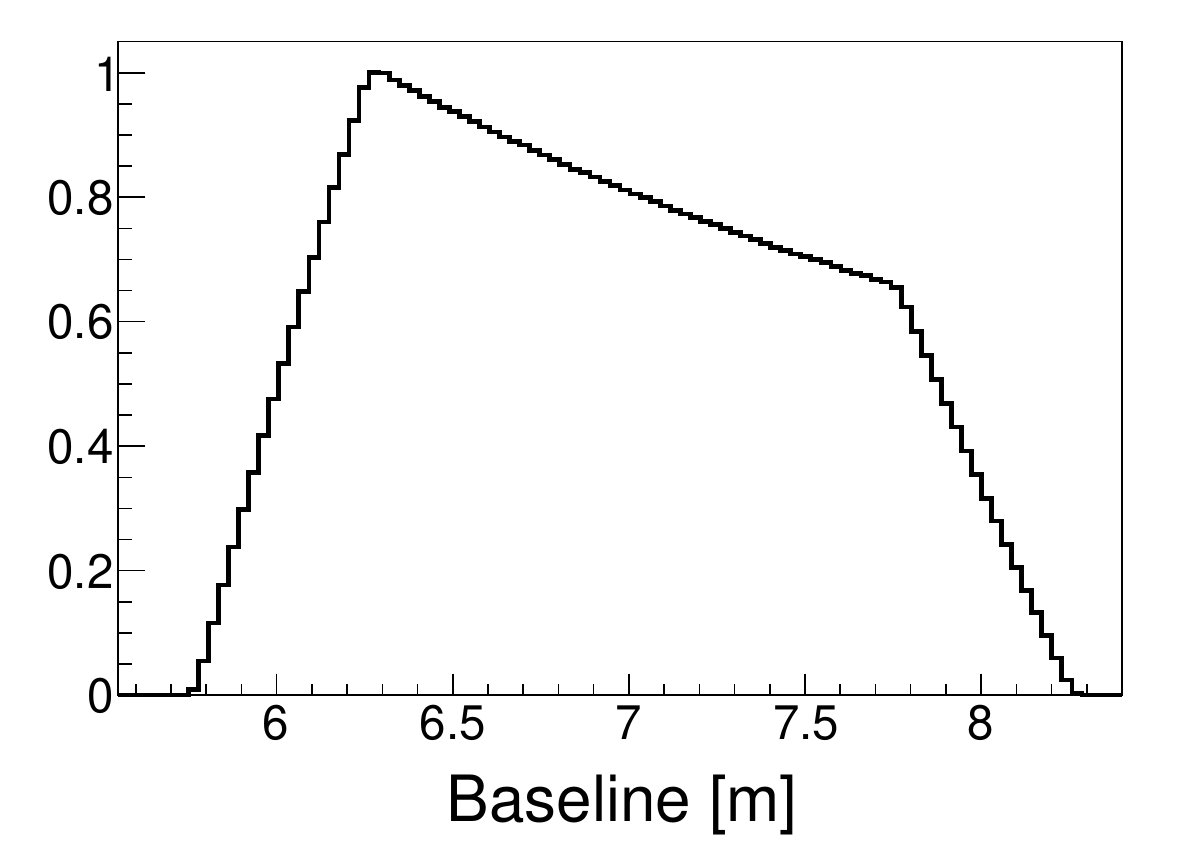}
  \caption{
    One-dimensional histograms of the generated pseudo-events. 
    \textbf{Left}: Neutrino energy spectrum sampled according to the $^{235}$U flux model by Huber~\cite{PhysRevC.84.024617} and the inverse beta decay cross-section by Vogel \& Beacom and Beacom~\cite{PhysRevD.60.053003}.
    \textbf{Center}: Distribution of baselines for accepted events at the HANARO site, assuming a 12-meter baseline and the detector configuration described in the text.
    \textbf{Right}: Baseline distribution for the Kijang research reactor site, based on a 7-meter baseline. 
    All distributions are based on $10^8$ generated events per site, taking into account detector geometry and spatial acceptance.}
  \label{fig:edist}
\end{figure}

In addition to spatial information, each event must be assigned a neutrino energy. As described in the previous section, neutrino energies are randomly sampled according to the $^{235}$U flux model by Huber~\cite{PhysRevC.84.024617}, while the inverse beta decay cross-section is calculated using the prescription by Vogel and Beacom~\cite{PhysRevD.60.053003}. Figure~\ref{fig:edist} displays the distribution of generated events as a function of neutrino energy and baseline.

\begin{figure}[t]
    \centering
    \includegraphics[width=0.48\textwidth]{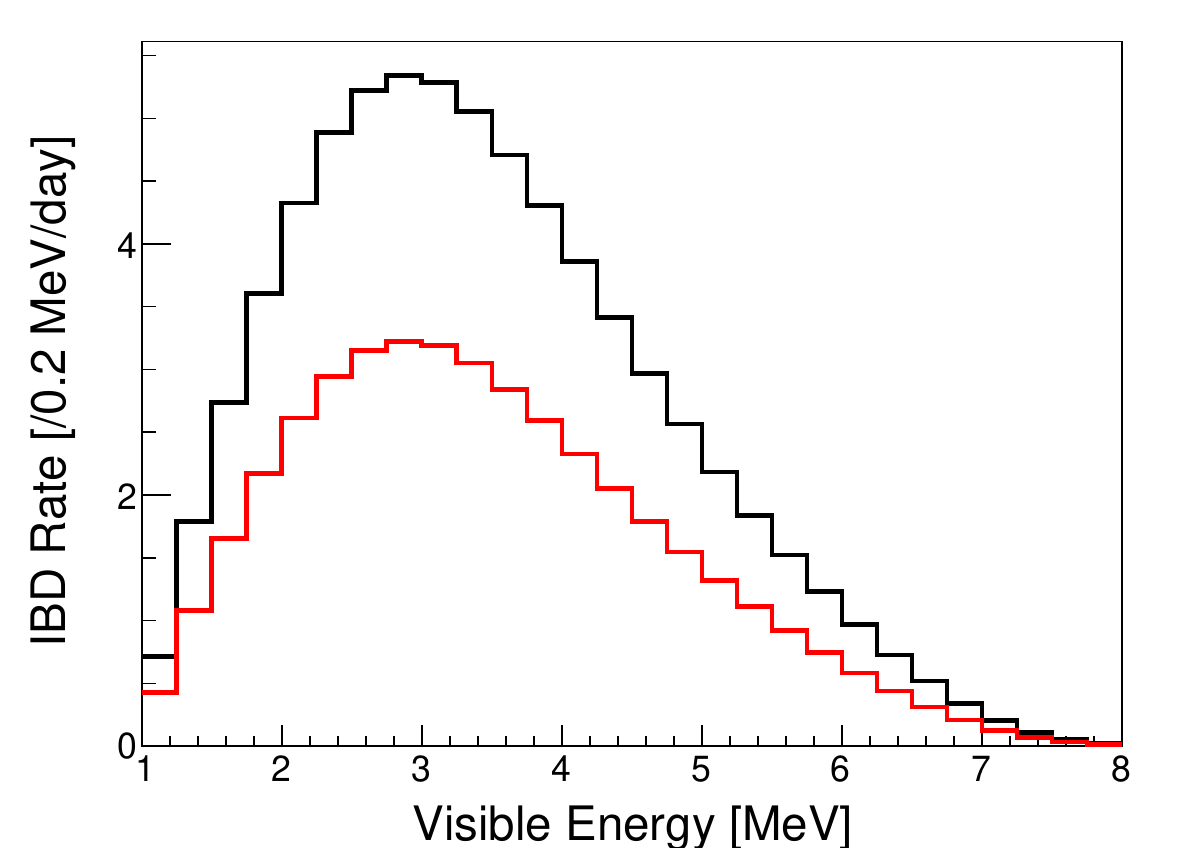}
    \includegraphics[width=0.48\textwidth]{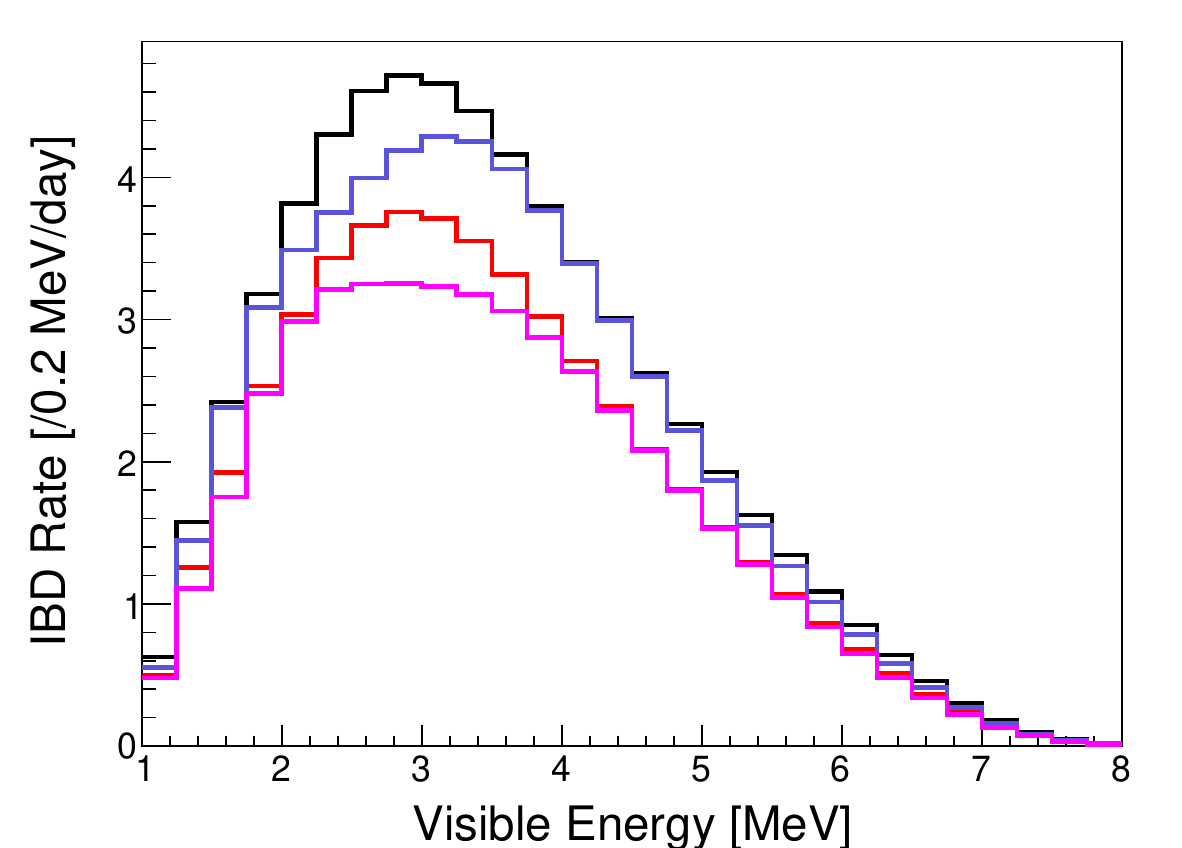}
    \caption{
        Simulated prompt energy spectra for selected detector segments.
        \textbf{Left:} Segment~1 energy distributions for the HANARO site (black) and Kijang site (red), both in the absence of oscillation.
        \textbf{Right:} Segment~2 and 4 energy spectra. Without oscillation, Segment~2 and 4 are shown in black and red, respectively. With oscillation ($\Delta m^2 = 2.4$~eV$^2$, $\sin^2 2\theta = 0.14$), Segment~2 and 4 appear in blue and magenta.
    }
    \label{fig:energy_distributions}
\end{figure}

The expected neutrino energy spectra, after being converted to prompt energies via the kinematics of the IBD reaction and smeared by the detector’s energy resolution, are shown in Figure~\ref{fig:energy_distributions}. Since the exact production location of the neutrino inside the reactor core is unknown, only the segment information of the interaction point within the detector is accessible. Consequently, I obtain the prompt energy spectra separately for each of the five detector segments, which are then used for oscillation analysis via segment-to-segment comparisons.

The number of expected IBD events per day is calculated according to the following expression:
\begin{equation}
N_{\mathrm{IBD}} = \frac{N_p \epsilon}{4\pi \langle L \rangle^2} \int \phi_{235}(E_\nu) \sigma_{\mathrm{IBD}}(E_\nu) \, dE_\nu \cdot \frac{P_{\mathrm{th}}}{E_{235}},
\end{equation}
where $N_p$ is the number of free protons in the liquid scintillator, and $\epsilon$ is the detection efficiency, assumed to be 50\%. $\langle L \rangle^2$ represents the average baseline, estimated from the generated event sample via $\langle 1/L^2 \rangle^{-1}$. The neutrino flux $\phi_{235}(E_\nu)$ and IBD cross-section $\sigma_{\mathrm{IBD}}(E_\nu)$ are taken from the Huber~\cite{PhysRevC.84.024617} and Vogel~\cite{PhysRevD.60.053003} models, respectively. The thermal power $P_{\mathrm{th}}$ uses the assumed values described in the previous section, and $E_{235} = 201.7$~MeV is the energy released per fission of $^{235}$U as reported in Ref.~\cite{PhysRev.181.1290}. Using this setup, the estimated number of expected IBD events per day is 187 for the HANARO site and 282 for the Kijang site.

Figure~\ref{fig:energy_distributions} (left) shows the prompt energy spectra at Segment~1, which is the closest to the reactor, for the HANARO site (black) and the Kijang site (red), both in the absence of oscillation. For the oscillated templates, 100 energy spectra were prepared assuming $\Delta m^2$ values ranging from 0.1 to 10\,eV$^2$. The right panel compares the spectra from Segment~2 and Segment~4 under both oscillated and non-oscillated scenarios. Without oscillation, the spectra for Segment~2 and 4 are shown in black and red, respectively, whereas with oscillation using the best-fit values of RAA~\cite{PhysRevD.83.073006}, the spectra are represented in blue and magenta, respectively.

As illustrated in the figure, when no oscillation is present, the energy spectra across all segments exhibit identical shapes, differing only in amplitude. However, due to the baseline dependence of neutrino oscillations, the inclusion of oscillation leads to distinct spectral shapes among the segments. This spectral distortion across segments is the foundation for the oscillation sensitivity of the experiment.

Now, I generate pseudo-data under the null hypothesis, assuming two years of experimental operation. The data includes not only the IBD signal but also backgrounds. The background is assumed to follow a $1/E^2$ shape, with 500 background events per day. The reactor-on data consist of the sum of the IBD signal and background, and the number of events in each energy bin are assumed to follow a Poisson distribution. The reactor-off data include only background events. Both reactor-on and reactor-off data are assumed to correspond to one year of measurement.

There are overall systematic uncertainties for the IBD signal and backgrounds, and the fluctuations due to these are included when generating the data set. The signal uncertainty accounts for the uncertainty that may occur in the number of protons in the target, the detection efficiency, and the baseline, and is conservatively set to 3\%. If it is analyzed as a ratio between modules, it is canceled out. Since the reactor-off data is used as the background, the difference between the off and on states is set as the uncertainty. The background associated with the reactor operation can vary greatly depending on the environment of the experimental site, but it is assumed to be absent in this study, and the overall uncertainty for the background is set to 10\%, the same as that used in NEOS.

Each pseudo-dataset consists of ten histograms: five segments for reactor-on and five for reactor-off, each representing the energy distribution. The analysis is based on the ratio of Segments 2 through 5 to Segment~1, using the difference between reactor-on and reactor-off datasets. The resulting ratio data are then compared to template spectra incorporating sterile neutrino oscillations.

To reflect these in the analysis, I construct the following chi-square formula:
\begin{equation}
    \chi^2=\sum_{i=2}^5\sum_{j=1}^{28}
    \frac{
        \left[D_{ij}/D_{1j} - E_{ij}(\Delta m^2,~\sin^2 2\theta)/
        E_{1j}(\Delta m^2,~\sin^2 2\theta)\right]^2
    }{
        \left(M_{ij}^\mathrm{ON}+\beta^2 M_{ij}^\mathrm{OFF}\right)/D_{1j}^2
        +\left(M_{1j}^\mathrm{ON}+\beta^2 M_{1j}^\mathrm{OFF}\right)\cdot D_{ij}^2/D_{1j}^4
    }
    +\frac{(\beta - 1)^2}{\sigma_b^2},
    \label{eq:chi2}
\end{equation}
where $i$ and $j$ represent the detector segment and energy bin, respectively. $M_{ij}$ denotes the number of the measured IBD events, while $E_{ij}$ is the number of the expected IBD events as a function of the oscillation parameters. $\beta$ is a nuisance parameter accounting for the overall uncertainty $\sigma_b$ in the background normalization, introduced to consider the difference between reactor-on and reactor-off data. The reactor-on data also include a 3\% uncertainty in the overall signal normalization, but this cancels out due to the ratio-based analysis. The difference data $D_{ij}$ are defined as:
\begin{equation}
    D_{ij}=M_{ij}^\mathrm{ON}-\beta M_{ij}^\mathrm{OFF}.
\end{equation}
For each mass square difference, the $\chi^2$ as a function of the oscillation amplitude $\sin^2 2\theta$ can be obtained from Eq.~\ref{eq:chi2}. By applying the function $P=\exp(-\Delta\chi^2/2)$, the corresponding probability density function~(PDF) can be derived.

\begin{figure}[t]
    \centering
    \includegraphics[width=0.48\textwidth]{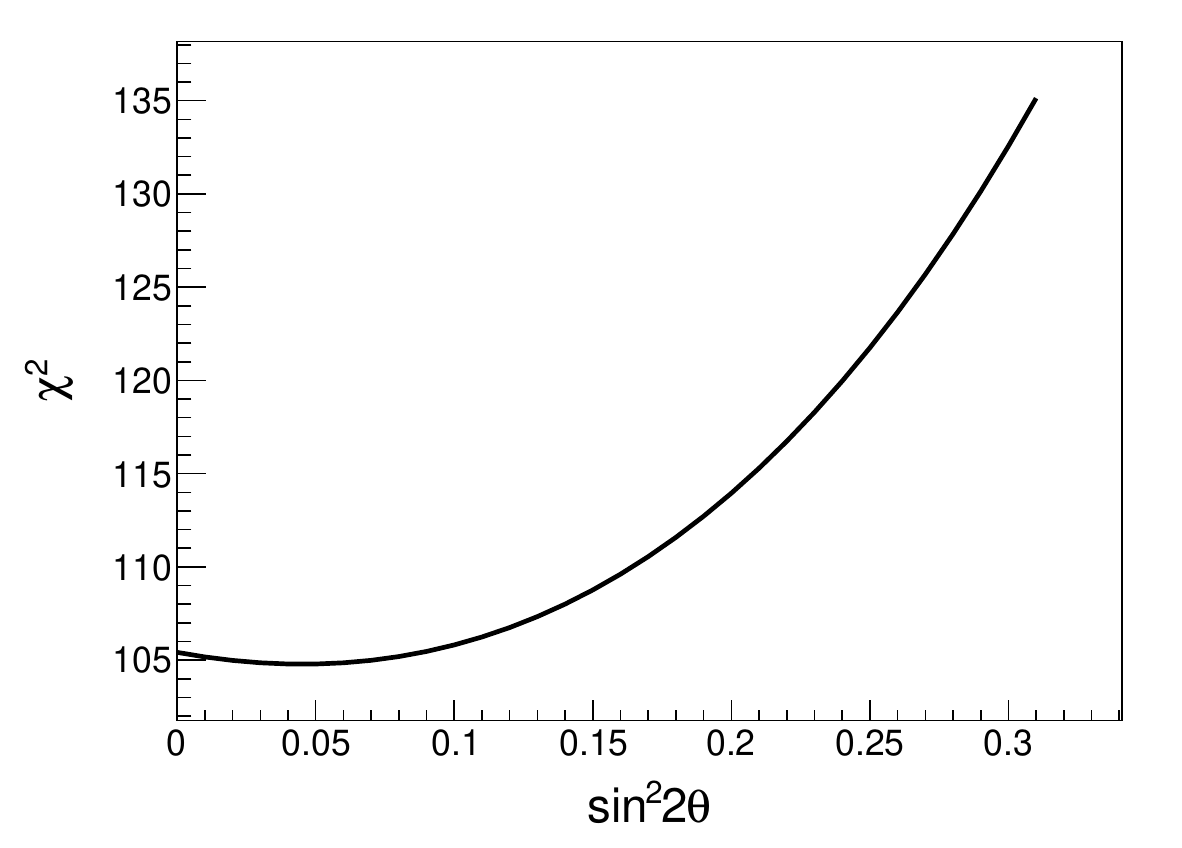}
    \includegraphics[width=0.48\textwidth]{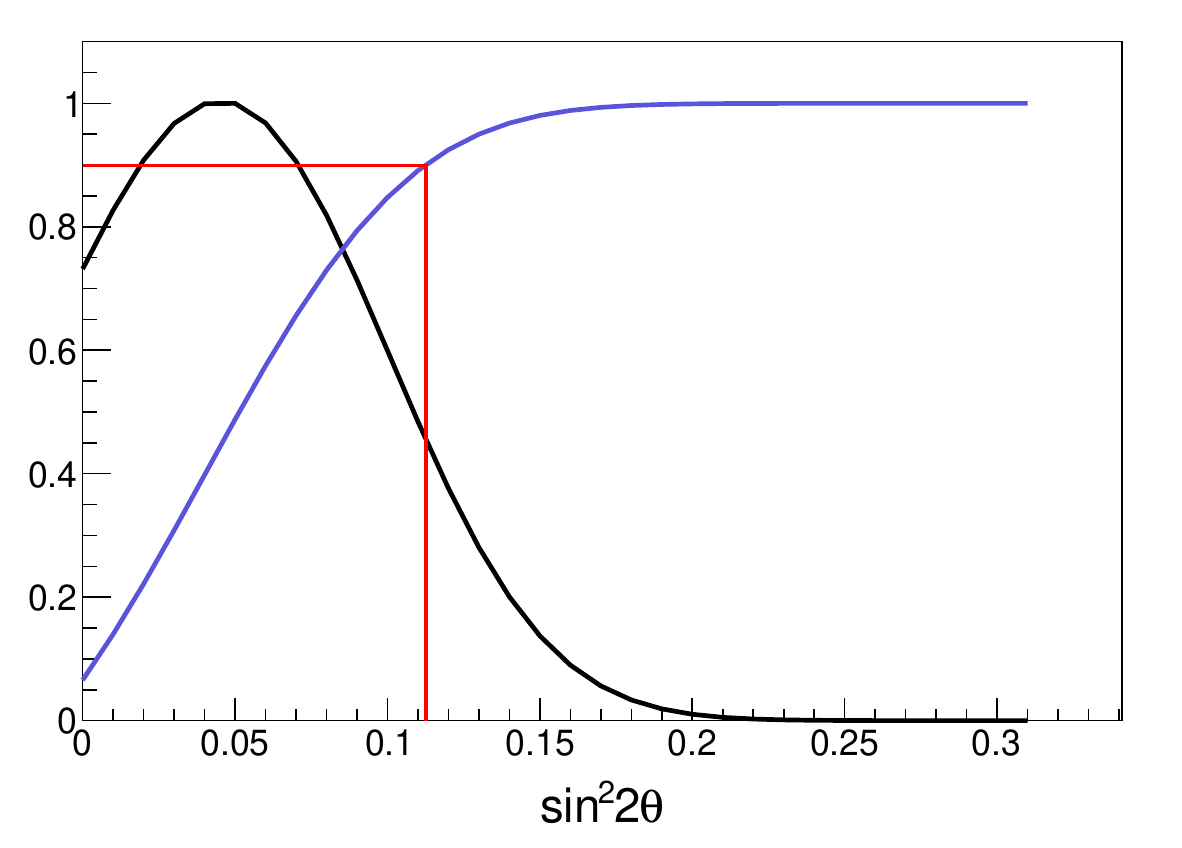}
    \caption{
        \textbf{Left:} The $\chi^2$ distribution as a function of the oscillation amplitude $\sin^2 2\theta$, evaluated for a fixed mass-squared difference ($\Delta m^2 = 1.02$\,eV$^2$) from one pseudo-dataset under the HANARO configuration. 
        \textbf{Right:} The corresponding probability density function (PDF; black) and cumulative distribution function (CDF; blue) derived from the $\chi^2$ curve. 
        The red line indicates the 90\% CL exclusion limit.
    }
    \label{fig:chi2_pdf_cdf}
\end{figure}

Figure~\ref{fig:chi2_pdf_cdf} presents the results of the analysis performed on a single pseudo-dataset generated under the HANARO setup. The left panel shows the $\chi^2$ values as a function of the oscillation amplitude $\sin^22\theta$, where the minimum indicates the best-fit value. The right panel displays the corresponding probability density function~(PDF) and cumulative distribution function~(CDF) derived from the $\chi^2$ curve using the relation $\exp(-\Delta\chi^2/2)$. The 90\% confidence level (CL) exclusion limit on $\sin^22\theta$ is determined by identifying the value for which the CDF reaches 0.9. The mean value of such exclusion limits obtained from 1000 pseudo-datasets defines as the sensitivity at the corresponding mass-squared difference.

\section{Sensitivity of Detector using Opaque LS}
For the sensitivity study of a detector using opaque liquid scintillator~(OLS), I reuse the same raw dataset mentioned in the previous section. This data set consists of 100 million simulated inverse beta decay~(IBD) events, each characterized by neutrino energy, baseline, and true IBD vertex within the detector. From these, I derive the prompt energy and apply position smearing to simulate the finite position resolution of the detector. These reconstructed quantities represent the observables in a realistic experimental scenario.

\begin{figure}[t]
  \centering
  \includegraphics[width=0.85\linewidth]{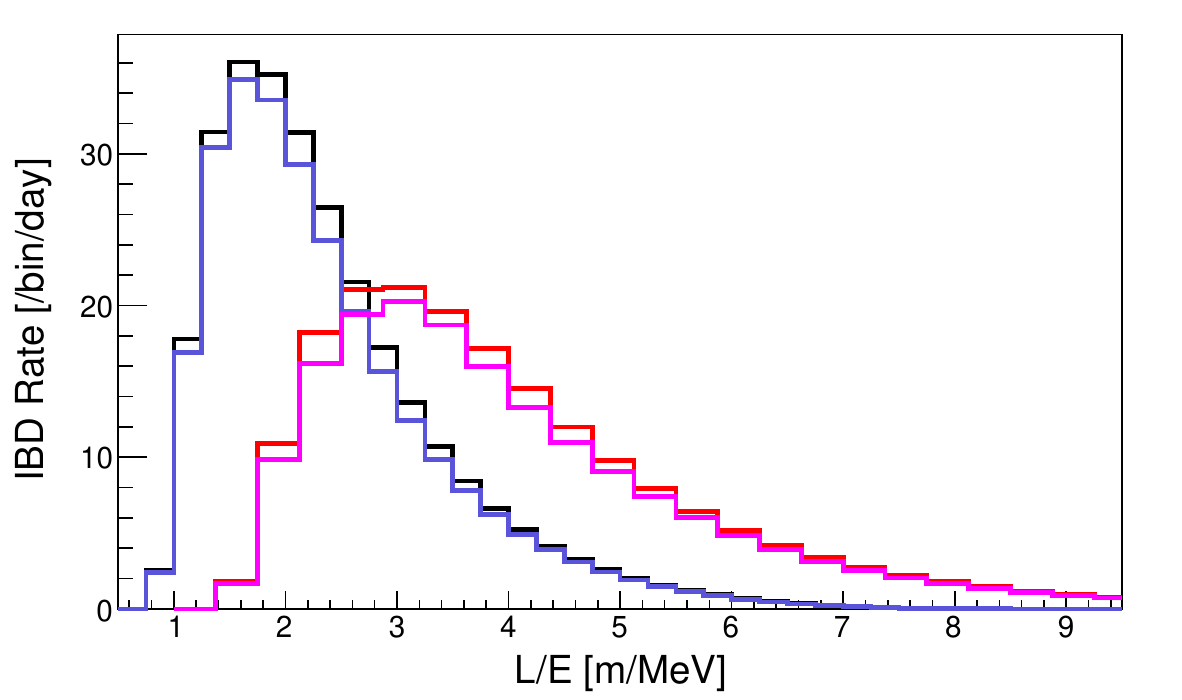}
  \caption{Templates of $L/E$ distributions for OLS detector simulations. The black and red lines show the no-oscillation hypothesis for HANARO and Kijang, while the blue and magenta lines correspond to an oscillation hypothesis with $\Delta m^2 = 2.4~\mathrm{eV}^2$ and $\sin^2 2\theta = 0.14$.}
  \label{fig:template_OLS}
\end{figure}

To construct the observable spectrum, I compute histograms of the $L/E$ distribution, where $L$ is the distance reconstructed from the center of the reactor core to the IBD interaction vertex, and $E$ is the reconstructed prompt energy. These $L/E$ histograms serve as templates both for generating pseudo-data and for performing oscillation fits. Figure~\ref{fig:template_OLS} shows the resulting $L/E$ distributions. The black and red lines represent the no-oscillation case for detectors located at HANARO and Kijang, respectively. The blue and magenta lines correspond to an oscillation scenario with the best-fit values of RAA, again for the HANARO and Kijang locations, respectively.

Assuming a one-year data-taking period with alternating reactor-on and reactor-off phases, I generate 1000 pseudo-experiments. Then each pseudo-dataset is analyzed using the $\chi^2$ fitting method. Unlike the segmented detector case that uses a spectral ratio analysis, the OLS detector uses an absolute $L/E$ spectrum analysis. Accordingly, the $\chi^2$ formula for the OLS case includes an overall normalization uncertainty of 3\% to account for systematic effects:
\begin{equation}
\chi^2 = \sum_{i=1}^{36}\frac{\left[M_i^\mathrm{ON}-\beta M_j^\mathrm{OFF}-\alpha E_i(\Delta m^2,\sin^22\theta)\right]^2}{M_i^\mathrm{ON}+\beta^2M_i^\mathrm{OFF}}
+\frac{(\alpha -1)^2}{\sigma^2}+\frac{(\beta - 1)^2}{\sigma_b^2},
\label{eq:chi2_OLS}
\end{equation}
where $\sigma$ is the uncertainty for overall normalization and $\alpha$ is the nuisance parameter corresponding to $\sigma$.

\begin{figure}[t]
  \centering
  \includegraphics[width=0.9\linewidth]{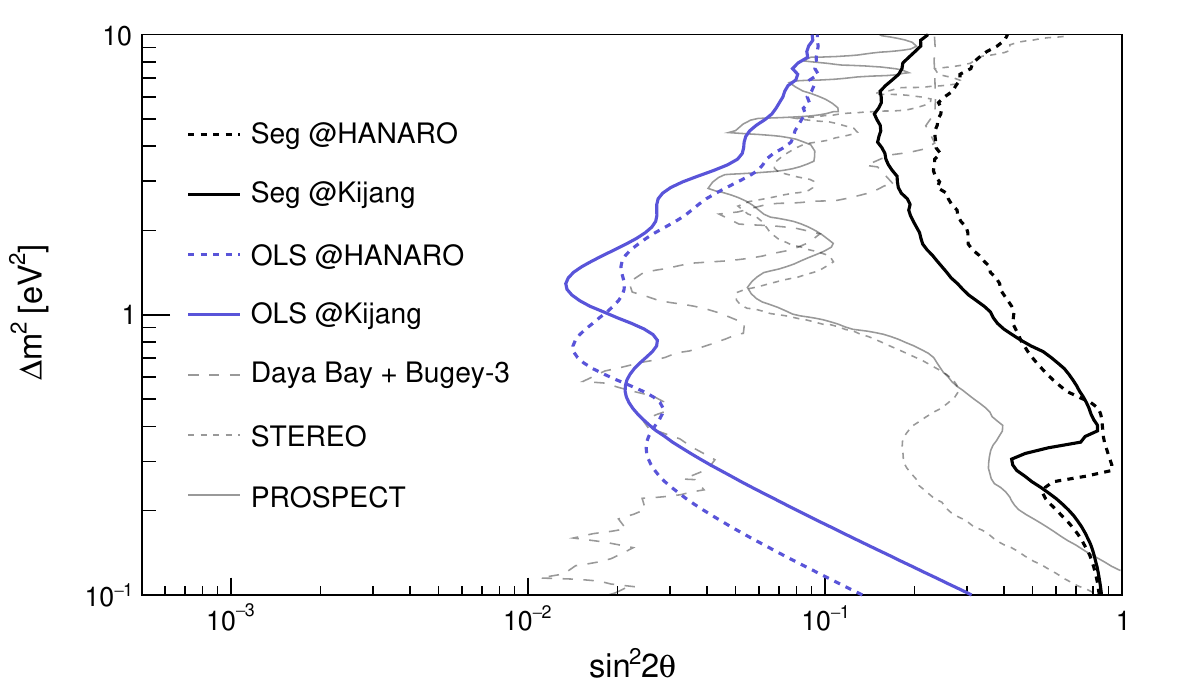}
  \caption{Comparison of detector sensitivities: segmented (black) and OLS (blue) detectors at HANARO (dashed) and Kijang (solid). Exclusion limits from Daya Bay + Bugey-3~\cite{PhysRevLett.117.151801}, STEREO~\cite{nature.613.257}, and PROSPECT~\cite{PhysRevD.103.032001} are overlaid for reference.}
  \label{fig:OLS_sensitivity}
\end{figure}

Using the resulting $\chi^2$ distributions as a function of oscillation parameters, I extract the 90\% confidence level exclusion limits for each pseudo-dataset. The average of these limits defines the sensitivity curve of the OLS detector. Figure~\ref{fig:OLS_sensitivity} compares the sensitivity results for four different configurations: segmented and OLS detectors located at HANARO and Kijang. The dashed lines indicate HANARO, and the solid lines indicate Kijang. The black and blue lines denote the segmented and OLS detector cases, respectively. Exclusion limits from existing experiments such as Daya Bay + Bugey-3, STEREO, and PROSPECT are also shown for comparison~\cite{PhysRevLett.117.151801,nature.613.257,PhysRevD.103.032001}.

As seen in Fig.~\ref{fig:OLS_sensitivity}, the segmented detector does not show a clear advantage in sensitivity under current assumptions, which include a conservative background and limited reactor power. In contrast, the OLS detector demonstrates world-leading sensitivity even under potentially challenging conditions. This superior performance arises from the excellent position resolution enabled by OLS technology, which significantly enhances the $L/E$-based oscillation analysis and allows a drastic reduction of background contamination.

\section{Conclusion}
In this work, I assessed the sensitivity of short-baseline reactor neutrino experiments to sterile neutrino oscillations by comparing two detector technologies: a conventional segmented scintillator detector and a novel opaque liquid scintillator (OLS) detector. Using a common simulation framework, I examined each detector type at two experimental sites, HANARO and Kijang, under realistic operational conditions.

The results indicate that, while the segmented detector approach is well-established, it offers limited advantages in the scenarios considered here, likely due to conservative estimates of background, limited reactor power, and modest detector performance. In contrast, the OLS detector exhibits significantly enhanced sensitivity, attributable to its excellent position resolution. This improvement enables precise reconstruction of the $L/E$ distribution and effective background suppression, which are critical for detecting subtle oscillation signals.

The OLS-based setup shows competitive or superior performance compared to current leading experiments such as PROSPECT, STEREO, and Daya Bay+Bugey-3, even in environments that are not optimized for neutrino experiments. This suggests that OLS technology, combined with compact reactor sources, could play a vital role in resolving the remaining tension in global sterile neutrino searches. Meanwhile, there are reports that Wilks' theorem may not apply to the analysis of sterile neutrinos, and there are attempts to interpret the positive signals of BEST and Neutrino-4 experiments based on this~\cite{Berryman2022,Coloma2021}. According to these, Wilks' theorem appears to strengthen the significance, but it may not make a big difference in relative comparison.

Future work may include the incorporation of detailed detector geometry, energy non-linearity effects, and more sophisticated background modeling. Additionally, experimental validation of the position reconstruction capability of OLS technology will be crucial to fully establish its potential in the context of precision neutrino physics.

\begin{acknowledgments}
This research was supported by the 2025 scientific promotion program funded by Jeju National University.
\end{acknowledgments}

\section*{Funding}
2025 scientific promotion program funded by Jeju National University.

\bibliography{reference}

\end{document}